\documentclass[a4paper,12pt]{article}
\usepackage{epsfig}
\usepackage[dvips,usenames]{color}
\usepackage{graphicx}

\newlength{\dinwidth}
\newlength{\dinmargin}
\begin{document}
\title{ Study of Semileptonic Decays $B^\pm \to \eta^{(\prime)} l \nu$ }

\bigskip

\author{ C.~S. Kim$^a$\footnote{cskim@mail.yonsei.ac.kr}
~~ and~~ Ya-Dong Yang$^b$\footnote{yang@physics.technion.ac.il}
\\
{$^a$ \small \it Department of Physics and IPAP, Yonsei University, Seoul 120-749, Korea}\\
{$^b$ \small \it Department of Physics, Technion-Israel Institute of Technology,
 Haifa 32000,  Israel}\\
}
\maketitle
\begin{picture}(0,0)
\put(305,290){\sf hep-ph/0107226}
\end{picture}

\bigskip\bigskip

\begin{abstract}
\noindent
We study semileptonic decays $B^\pm \to \eta^{(\prime)} l \nu$, which are 
suggested to be used to extract the hadronic form factors of  $B$ meson decays to 
$\eta(\eta^{\prime})$ and  the angle of $\eta-\eta^{\prime}$ mixing. 
This would be of great benefit to theoretical studies of $B$ nonleptonic decays involving
$\eta$ and $\eta^{\prime}$, and could lead to
a reliable and complementary determination of  $V_{ub}$. 
The branching ratios are estimated to be ${\cal B}(B^\pm \to\eta^{(\prime)} l \nu)=
4.32 \pm 0.83~(2.10 \pm 0.40) \times 10^{-5}$,
which could be extensively studied experimentally at  $BaBar$ and Belle.
\end{abstract}

\newpage

Semileptonic $B$ decays are  subjects of considerable interests that have been 
extensively studied with applications of various nonperturbative theorectical  
frameworks. They offer the most direct method to determine the weak mixing angles and 
to probe strong interaction confinement phenomenology of hadronic transitions.
Recently $V_{cb}$ has been determined from semileptonic $B$ decays and becomes 
the third most accurately measured Cabibbo-Kobayashi-Maskawa (CKM)
matrix element \cite{pdg}. CLEO Collaborations \cite{cleo} have made  measurments of 
the decays $B^0 \to \pi^- l^+ \nu$ and $\rho^{-} l^{+} \nu$ with the results
\begin{eqnarray}
{\cal B}(B^0 \to \pi^- l^+ \nu) &=& (1.8\pm 0.4\pm 0.3\pm 0.2)\times10^{-4}, \nonumber\\
{\cal B}(B^0 \to \rho^- l^+ \nu) &=& (2.57\pm 0.29^{+0.33}_{-0.46}\pm 0.41)\times
       10^{-4} \nonumber\\
{\rm and} ~~~ |V_{ub}|&=& (3.25\pm0.14^{+0.21}_{-0.29}\pm 0.55)\times10^{-3}.\nonumber
\end{eqnarray} 
It is known that extracting $|V_{ub}|$ from the measured decay rates require 
significant inputs 
from theoretical estimations of the hadronic form factors which involve complex 
strong-interaction dynamics. 
With $BaBar$ and Belle  taking data, we are entering a new era of $B$ physics. 
Prospects for accurate measurement of these decay modes  become excellent. We can  
foresee that the decays $B^\pm \to \eta l \nu$ and $\eta^{\prime}l\nu$ could be also 
observed at $B$ factories in the near future. In this Brief Report, we study 
the decays $B^\pm \to \eta l \nu$ and $\eta^{\prime}l\nu$  to show that many 
interesting physical observables can be extracted from measurments of these decays.   
   
Amplitudes of exclusive semileptonic $B \to P l \nu$ 
($l=\mu,\,e$ and $P=\pi, \eta, \eta^{\prime}$ )
can be 
written as 
\begin{equation}
{\cal M}(B\to P l \nu)=\frac{G_{F}}{\sqrt 2}V_{ub}~ \bar{l}\gamma_{\mu}(1-\gamma_{5})\nu ~
\langle P(p_{P})|\bar{u}\gamma^{\mu}(1-\gamma_{5})b|B(p_{B})\rangle,
\end{equation}
where the hadronic transition matrix can be parameterized as
\begin{equation}
\langle P(p_{P})|\bar{u}\gamma_{\mu}(1-\gamma_{5})b|B(p_{B})\rangle=
F_{+}^{B\to P}(q^{2}) (p_{B}+p_{P})_{\mu}+F_{-}^{B\to P}(q^{2}) (p_{B}-p_{P})_{\mu} \ .
\end{equation}  
Here, $q=p_{B}-p_{P}$, and $F_{+(-)}^{B\to P}(q^{2}) $ are the relevant form factors. 
Using these notations, the double differential decay width is 
\begin{equation} 
\frac{d\Gamma(B\to P l\nu)}{dE_{l}dq^2}=G^{2}_{F}|V_{ub}|^2 \frac{1}{16\pi^3 M_{B}}
\left|F^{B\to P}_{+}(q^{2})\right|^2 
\left[ 2E_{l}( m^{2}_{B}+q^2 -m^2_{P})-m_{B}(4 E^2_{l}+q^2 ) \right],
\end{equation}
where we have neglected the lepton mass.

To calculate the semileptonic decay width, we have to know precisely the form factors 
$F^{B\to P}_{+}(q^2 )$, which challenge our poor knowledge of nonperturbative QCD. 
In recent years, considerable progress has been made in the calculations of 
$F^{B\to\pi}_{+} (q^{2})$ with various theoretical approaches:  quark models
\cite{qm}, QCD sum rules \cite{ruckl, qcdsr} and lattice QCD \cite{ukqcd,lattice}.
Combining the results of different approaches, say, predictions of QCD sum rules 
in low $q^2$ region and of lattice QCD in high $q^2$ region,  
we could  possibly obtain a good theoretical description of $F^{B\to\pi}_{+} (q^{2})$ in
the whole $q^2$ region. However, both QCD sum rule and lattice calculations of the form factors
$F^{B\to \eta^{(\prime)}}_{+}(q^2 )$ are not yet available in the literature. 
Therefore, we will use 
$SU(3)_F$ symmetry to relate them to $F^{B\to\pi}_{+} (q^{2})$. For $\eta-\eta^{\prime}$ 
mixing, we adopt the scheme \cite{rosner, FKS, scadron}
 \begin{eqnarray}
|\eta\rangle&=& \cos\phi|\eta_{q}\rangle-\sin\phi|\eta_{s}\rangle, \nonumber \\
|\eta^{\prime}\rangle &=& \sin\phi|\eta_{q}\rangle+\cos\phi|\eta_{s}\rangle,
\end{eqnarray}
where $|\eta_{q}\rangle=(u\bar{u}+d\bar{d})/{\sqrt 2}$, 
$|\eta_s \rangle=s\bar{s}$, and $\phi=39.3^{\circ}$ is the fitted
mixing angle \cite{FKS}. 
Assuming $SU(3)_F$ symmetry, the form factors $F^{B\to\eta^{(\prime)}}_{+}(q^2 )$ are 
related to  $F^{B\to\pi}_{+}(q^2 )$ by the relations
\begin{equation}
F^{B\to\eta}_{+} (q^2 ) = \cos\phi \; F^{B\to\pi}_{+}(q^{2}),~~~~~ 
F^{B\to\eta^{\prime}}_{+} (q^2 ) = \sin\phi \; F^{B\to\pi}_{+}(q^{2}).
\end{equation} 
The  form factor $F^{B\to\pi}_{+}(q^{2})$  is known to 
be dominated by $B^{\ast}$ pole in the small--recoil region $q^2 \sim (m_B -m_{\pi})^2$ and to
scale as $F^{B\to\pi}_{+}(q^{2}\simeq m^2_{B})\sim \sqrt{m_{B}}$ in the heavy quark limit
\cite{IW}. 
Recent studies \cite{lattice, charles, cheng} imply the dipole 
behavior for $F_{+}(q^{2})$ in the large--recoil region  $q^2 \sim 0$.  
The easiest way to extrapolate the $q^2$ dependence is to suppose the dipole 
behavior for  $F_{+}(q^{2})$ \cite{ukqcd, alexan}
\begin{equation}
F_{+}^{B\to\pi}(q^2 )=\frac{F_{+}^{B\to\pi}(0)}{(1-\frac{q^2}{m^2_{B^*}})^2}, 
~~~~ ({\rm Dipole}) 
\end{equation}
where $m_{B^*}$ is the pole mass of $B^{*}(1^- )$ associated with the weak current
induced by the decay. 

Becirevic and Kaidalov (BK) \cite{BK} have also proposed 
a numerical  parameterization which satisfies the heavy quark scaling laws \cite{IW}
and most of the known constraints \cite{BK},
 \begin{equation}
F_{+}^{B\to\pi}(q^2 )=\frac{c_{B}(1-\alpha_{B\pi})}
{(1-\frac{q^2}{m^2_{B^*}})(1-\alpha_{B\pi} \frac{q^2}
{m^2_{B^*}}  ) }~.
\end{equation}
We can read from here $F_{+}^{B\to\pi}(0)=c_{B}(1-\alpha_{B\pi})$. 
Using BK parameterization to fit their light--cone QCD  sum rule (LCSR) calculations,
Khodjamirian $et\, al$. found 
$\alpha_{B\pi}=0.32^{+0.21}_{-0.07}$ and $F_{+}^{B\to \pi}(0)=0.28\pm0.05$ \cite{ruckl}.
The recent results from Lattice QCD are \cite{Roma}
\begin{eqnarray}
\alpha_{B\pi}=0.40\pm0.15,&F_{+}^{B\to \pi}(0)=0.26\pm0.05, ~~~~~~({\rm Lattice~I})\nonumber\\
\alpha_{B\pi}=0.45\pm0.17,&F_{+}^{B\to \pi}(0)=0.28\pm0.06, ~~~~~~({\rm Lattice~II})
\end{eqnarray}
where the two sets of results (Lattice~I, II) correspond to two different methods used
in Ref. \cite{Roma}.

To eliminate the effect of large uncertainty in $V_{ub}$,
we  relate the branching ratios  
$B\to \eta^{(\prime)}l\nu$  to ${\cal B}(B^- \to\pi^0 l\nu )$, and  get 
\begin{eqnarray}
{\cal R}_1 = \frac{{\cal B}(B^- \to\eta l\nu)}{{\cal B}(B^{-}\to \pi^0 l \nu)}
&=& |\cos\phi|^{2} \frac{\int^{(m_{B}-m_{\eta})^2}_{0}dq^2 |F_{+}^{B\to \pi}(q^{2})|^2 
\left( (m^2_{B}+m^2_{\eta}-q^2 )^2 -4 m^2_{B}m^2_{\eta}\right)^{\frac{3}{2}}  }
{ \int^{(m_{B}-m_{\pi})^2}_{0}dq^2 |F_{+}^{B\to \pi}(q^{2})|^2 
\left( (m^2_{B}+m^2_{\pi}-q^2 )^2 -4 m^2_{B}m^2_{\pi}\right)^{\frac{3}{2}}}\nonumber\\
&=&|\cos\phi|^{2} \times  \left\{ 
\begin{array}{cccc}
0.527, & ({\rm Dipole})\\
0.813, & ({\rm LCSR})\\
0.802, & ({\rm Lattice~ I})\\
0.794, & ({\rm Lattice~ II})
\end{array} \right.
\\
 {\cal R}_2 = \frac{{\cal B}(B^- \to\eta' l\nu)}{{\cal B}(B^{-}\to \pi^0 l \nu)}
&=& |\sin\phi|^{2} \frac{\int^{(m_{B}-m_{\eta'})^2}_{0}dq^2 |F_{+}^{B\to \pi}(q^{2})|^2 
\left( (m^2_{B}+m^2_{\eta'}-q^2 )^2 -4 m^2_{B}m^2_{\eta'}\right)^{\frac{3}{2}}  }
{ \int^{(m_{B}-m_{\pi})^2}_{0}dq^2 |F_{+}^{B\to \pi}(q^{2})|^2 
\left( (m^2_{B}+m^2_{\pi}-q^2 )^2 -4 m^2_{B}m^2_{\pi}\right)^{\frac{3}{2}}}\nonumber\\
&=&|\sin\phi|^{2} \times
\left\{ 
\begin{array}{cccc}
0.310, & ({\rm Dipole})\\
0.599, & ({\rm LCSR})\\
0.584, & ({\rm Lattice~ I})\\
o.573, & ({\rm Lattice~ II}). 
\end{array} \right.
\end{eqnarray}
Using the CLEO reslut \cite{cleo}, ${\cal B}(B^{0}\to \pi^+ l \nu)=(1.8\pm0.6)\times10^{-4}$, 
and the relations ${\cal B}(B^{0}\to \pi^+ l \nu)=2 {\cal B}(B^{-}\to \pi^0 l \nu)$,   
$\phi=39.3^{\circ}$, we get  
\begin{eqnarray}
{\cal B}(B^- \to\eta l\nu)&=& \left\{ 
\begin{array}{cccc}
(2.84 \pm 0.95) \times 10^{-5}, & ({\rm Dipole})\\
(4.38 \pm 1.46) \times 10^{-5}, & ({\rm LCSR})\\
(4.32 \pm 1.44) \times 10^{-5}, & ({\rm Lattice~ I})\\
(4.28 \pm 1.42) \times 10^{-5}, & ({\rm Lattice~ II}),
\end{array} \right.
\\
{\cal B}(B^- \to\eta^{\prime} l\nu)&=&\left\{ 
\begin{array}{cccc}
(1.12 \pm 0.37) \times 10^{-5}, & ({\rm Dipole})\\
(2.16 \pm 0.72) \times 10^{-5}, & ({\rm LCSR})\\
(2.10 \pm 0.70) \times 10^{-5}, & ({\rm Lattice~ I})\\
(2.06 \pm 0.68) \times 10^{-5}, & ({\rm Lattice~ II}).
\end{array} \right.
\end{eqnarray}
We can see that the predictions of LCSR form factors \cite{ruckl} agree very well
with those of Lattice (I, II) QCD \cite{Roma}. Averging  predictions from Lattice QCD and
LCSR,   we obtain
\begin{eqnarray}
{\cal B}(B^- \to\eta l\nu)&=& (4.32 \pm 0.83) \times 10^{-5} , \nonumber \\
{\cal B}(B^- \to\eta^{\prime} l\nu)&=& (2.10 \pm 0.40) \times 10^{-5}  .
\end{eqnarray}

We note that the  ratios ${\cal R}_1$ and ${\cal R}_2$  are  independent of the 
value of $F^{B\to\pi}_{+}(0)$,  but very sensitive to the details of its $q^2$ dependence. 
We also note that to give the same numerical  predictions  for $B\to \pi l\nu$, 
the $F^{B\to\pi}_{+}(0)$ for 
dipole parameterization should be smaller than that for BK parameterization. If the same 
value  $F^{B\to\pi}_{+}(0)$ is used in both BK and dipole parameterizations, one will find    
\begin{equation}
{\cal R}_3 = \frac{{\cal B}^{{\rm dipole}}(B^- \to\pi^0 l\nu)}
     {{\cal B}^{{\rm LCSR}}(B^{-}\to \pi^0 l \nu)} =3.13,  
\end{equation}
which implies that the dipole form factor will overestimate the decay rates
because $\pi$ meson is very light and the lepton pair invariant mass 
$q^2$ can be very near the $B^*_u$ pole. Therefore, 
theoretical predictions for $B\to\pi l\nu$ (and $B\to\eta^{(\prime)} l\nu$ in turn)  
are very sensitive 
to the $q^2$ dependence of $F_{+}^{B\to\pi}(q^2)$. 
It is well known that the extraction of $V_{ub}$ from decay rates of 
$B\to \pi (\rho) l \nu$ suffers from large  theoretical uncertainities in the hadronic form factors. 
Testing the predictions and eventual measurements of $d \Gamma/dq^2$ can
provide valuable information on the hadronic form factors governing
$b \to u l \nu$ decays, and hence lead to a reliable determination of $V_{ub}$.
With much more data to arrive soon from $B$ factories, 
the $q^2$ and the lepton energy distributions  can be precisely measured and be
used to distinguish these form factor parameterizations, and to extract  $V_{ub}$.
The determination of $V_{ub}$ from $B \to \eta^{(\prime)} l \nu$ would 
represent a powerful method complementary
to the determination of $V_{ub}$ from other exclusive decay modes,
e.g., from $B \to \pi (\rho) l \nu$.

In Fig. 1,
we plot  the $q^2$ distributions and the lepton energy $E_l$ distributions of the 
decays $B^{-}\to\eta^{(\prime)} l \nu$, where we have normlized the form factors to 
give ${\cal {\cal B}}(B^{-}\to\pi^0 l \nu)=9\times 10^{-5}$. We find that both LCSR and
Lattice QCD predict very consistent lepton energy distributions as well as
consistent decay rates for the decays.
Integrating out the lepton energy in Eq. (3), one obtains
\begin{equation} 
\frac{d\Gamma(B\to P l\nu)}{dq^2}=\frac{ G^{2}_{F}|V_{ub}|^2 }{192\pi^3 M^3_{B}}
\left|F^{B\to P}_{+}(q^{2})\right|^2 \left[ ( m^{2}_{B}+m^2_{P}-q^2  )^{2}
-4 m^{2}_{B}m^2_{P}\right]^{\frac{3}{2}}.
\end{equation}
At maximum recoil point $(q^2 =0)$, we have 
\begin{eqnarray}
{\cal R}_4 =\left. \frac{d\Gamma(B\to \eta^{(\prime)} l\nu)/dq^2}{d\Gamma(B^- \to\pi^{0} l\nu)/dq^2}
\right|_{q^2 =0}
 &=&\frac{(m^2_B -m^2_{\eta^{(\prime)}})^3 }{(m^2_B -m^2_{\pi})^3 }
\left| \frac{F^{B\to\eta^{(\prime)} }_+ (0)}{F^{B\to\pi^0 }_+ (0)}\right|^2 ,\\
{\cal R}_5 =\left. \frac{d\Gamma(B\to \eta^{\prime} l\nu)/dq^2}{d\Gamma(B^- \to\eta l\nu)/dq^2}
\right|_{q^2 =0}
 &=& \frac{(m^2_B -m^2_{\eta^{\prime}})^3 }{(m^2_B -m^2_{\eta})^3 }
\left| \frac{F^{B\to\eta^{(\prime)} }_+ (0)}{F^{B\to\eta }_+ (0)}\right|^2 \\
   &=&
\frac{(m^2_B -m^2_{\eta^{\prime}})^3 }{(m^2_B -m^2_{\eta})^3 }
\left| \cot\phi \right|^2 . \nonumber
\end{eqnarray}
As indicated by QCD sum rule calculations \cite{ruckl,qcdsr}, the value of
$F^{B\to\eta^{(\prime)} }_+ (q^2 )$ is rather stable 
under the variation of
$q^2$ when the value of $q^2$ is small. So 
the ratios ${\cal R}_4 $  and ${\cal R}_5 $ can be safely extrapolated to few GeV$^2$ to 
make the phase spaces sizable.   
Once the ratios are measured, they can be used to extract the form factor 
$F^{B\to \eta^{(\prime)} }_+ (0)$ and the mixing angle $\phi$ from the above relations.

In the literature, semileptonic decays $D_s \to \eta(\eta')l\nu$ have been taken as
sources of extracting $\eta-\eta'$ mixing angle and testing the mixing
schemes \cite{FKS, melikhov}. We note that the decays  $D_s \to \eta(\eta')l\nu$
involve strange contents $|\eta_s \rangle$ of $\eta(\eta')$, and  $B^- \to \eta(\eta')l\nu$
involve non-strange contents $|\eta_q \rangle$ of $\eta(\eta')$, therefore 
 $B^- \to \eta(\eta')l\nu$ and  $D_s \to \eta(\eta')l\nu$ could provide combined testing 
of $\eta-\eta'$ mixing scheme. 

As it is well
known that $\eta$ and $\eta' $ are too complicated objects to be reliably described
within QCD yet, 
it may be very hard to calculate the transition form factors 
$F^{B\to\eta^{(\prime)}}_{+}(q^2)$ 
within the frameworks of lattice QCD and QCD sum rules. 
The experimental extraction
of those form factors will improve our theorectical understanding of many interesting 
nonleptonic $B$ decay modes involving $\eta$ and $\eta^{\prime}$, and might shed light on
the problem, currently under discussion \cite{Du}, of the puzzling large branching ratios of 
$B \to K\eta'$ observed by CLEO \cite{CLEOeta}.

Finally, we note a few experimental comments:
Background for $B \to \eta^{(\prime)} l \nu$ would be much smaller than
that of
$B \to \pi l \nu$, due to much lower multiplicity, since the random background caused by 
$B \to \eta X$ is about an order of magnitude smaller than that by $B \to \pi X$.
The reconstruction of $\eta \to \gamma \gamma$ in experimental analyses
may be much easier than $\pi^0 \to \gamma \gamma$, 
even though the signal/noise ratio is worse, 
because the mass of $\eta$ is much bigger than that of $\pi^0$.
And we could even require the momentum of $\eta$
to be bigger than 1 GeV to remove  combinatorial backgrounds substantially.

To conclude, we studied semileptonic decays $B^\pm \to \eta^{(\prime)} l \nu$, which can 
be used to extract the hadronic form factors of  $B$ meson decays to 
$\eta(\eta^{\prime})$ and  the angle of $\eta-\eta^{\prime}$ mixing. 
The branching ratios are found to be 
${\cal B}(B^- \to\eta^{(\prime)} l \nu)=4.32 \pm 0.83~(2.10 \pm 0.40)\times 10^{-5}$.


\section*{Acknowledgments}

We thank G. Cvetic, Hongjoo Kim and Y.J. Kwon for careful reading of the manuscript and 
their valuable comments.
The work of C.S.K. was supported 
by Grant No. 2000-015-DP0077 of the KRF.
Y.Y  is supported by the US-Israel Binational Science
Foundation and the Israel Science Foundation.


\newpage
\vskip 30mm

\begin{figure}[htbp] 
\scalebox{1}{\epsfig{file=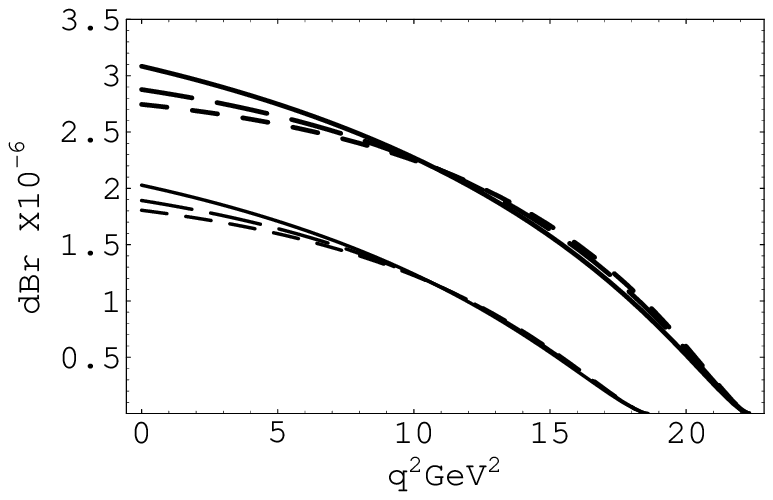}}
\scalebox{1}{\epsfig{file=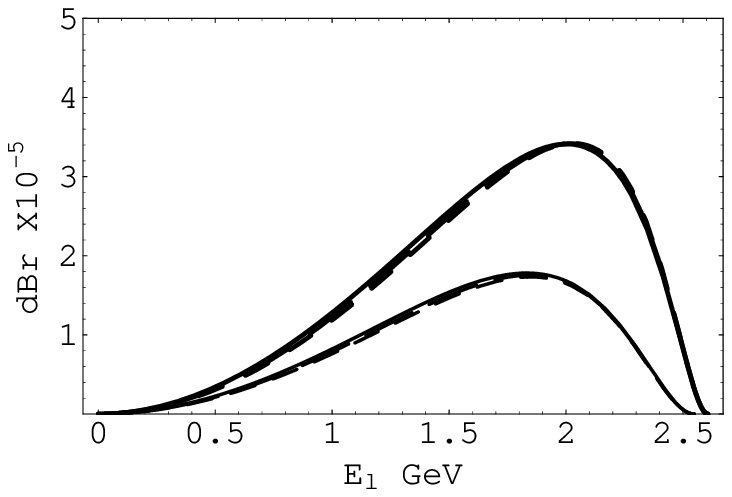}}
\caption{\small The spectra $d{\cal B}/dq^2$ as function of $q^2$  and the 
spectra $d{\cal B}/dE_l$ as function of the electron energy $E_l$. The thick solid, 
long-dashed and  short-dashed  
curves are the distributions of $d{\cal B}(B\to\eta l\nu)$ with LCSR \cite{ruckl} and
Lattice (I, II) QCD \cite{Roma} form factors,
and the thin  curves are those for $d{\cal B}(B\to\eta' l\nu)$. 
}  
\end{figure} 

\end{document}